\documentclass{article}
\usepackage{spconf,amsmath,graphicx}
\usepackage{multirow}
\usepackage{makecell}

\title{DFSMN-SAN with Persistent Memory Model for Automatic Speech Recognition}
\name{Zhao You$^1$, Dan Su$^1$, Jie Chen$^1$, Chao Weng$^2$, Dong Yu$^2$}
\address{$^1$Tencent AI Lab, Shenzhen, China\\
$^2$Tencent AI Lab, Bellevue, WA, USA \\
\{dennisyou, dansu, leojiechen, cweng, dyu\}@tencent.com}
\begin{document}
%
\maketitle
\begin{abstract}

Self-attention networks (SAN) have been introduced into automatic speech recognition (ASR) and achieved state-of-the-art performance owing to its superior ability in capturing long term dependency. One of the key ingredients is the self-attention mechanism which can be effectively performed on the whole utterance level. In this paper, we try to investigate whether even more information beyond the whole utterance level can be exploited and beneficial.
We propose to apply self-attention layer with augmented memory to ASR. Specifically, we first propose a variant model architecture which combines deep feed-forward sequential memory network (DFSMN) with self-attention layers to form a better baseline model compared with a purely self-attention network. Then, we propose and compare two kinds of additional memory structures added into self-attention layers.
Experiments on large-scale LVCSR tasks show that on four individual test sets, the DFSMN-SAN architecture outperforms vanilla SAN encoder by 5\% 
relatively in character error rate (CER).
More importantly,  the additional memory structure provides further 5\% to 11\% relative improvement in CER.

\end{abstract}
\vspace{-15pt}
\section{Introduction}
\label{sec:intro}


Transformer models \cite{vaswani2017attention} have been successfully applied and proven to be more effective than recurrent neural networks, eg., LSTMs, in several NLP tasks.  
The two key ingredients include sinusoidal positional encoding and the self-attention mechanism to be context-aware on input word embeddings. Recently, transformer models and their variants have also been actively investigated for speech recognition as well \cite{povey2018time, sperber2018self, chiu2018state, dong2018speech}. To work well for ASR modeling, transformer architecture needs to make some revision. Some key points have been summarized by previous work.

First, due to the fact that a speech utterance typically lasts for couple of seconds, while speech frames are extracted using a several-milliseconds window, downsampling acoustic input frames to widen temporal context is always beneficial. Some variant of model architecture has also been explored either using TDNN \cite{dong2018speech} or BiLstm \cite{sperber2018self} as front blocks to extract high-level features for the self-attention layers. Second, transformer encoder is effective both with CTC loss \cite{graves2006connectionist} and within Listen, Attend and Spell (LAS) framework. In \cite{salazar2019self}, the CTC loss  was applied to optimize the Transformer encoder structure for ASR. In \cite{zhou2018syllable, dong2018speech}, the entire encoder-decoder structure of the original Transformer was examined in the context of Mandarin Chinese speech recognition tasks. 
Last but not least, with self-attention network, a common observation is that the longer the context is selected, the better the performance can be obtained. In \cite{dong2019self}, to support online speech recognition, a chunk-hopping mechanism is proposed. The experimental results showed that compared with the whole utterance context,  chunk-hopping always leads to performance degradation. 
In this work, the priority is given to explore all possible information to improve speech recognition accuracy, while streaming speech recognition where latency is an important factor to consider will not be in the scope of the study.

To make further improvement based on self-attention layer, a natural idea is to explore more information beyond the whole utterance context length. Recently, in \cite{Sainbayar2019augmenting}, the authors proposed to augment self-attention with persistent memory by introducing a new layer that merges the self-attention and feed-forward sub-layers into a single unified attention layer. It has shown that the additional persistent memory block in the form of key-value vectors stores some global information that does not depend on the context. This work sheds light on achieving our above motivation.

In this paper, we first explore a variant of model architecture which combines DFSMN and self-attention layers. Experiments show this model architecture outperforms a standard transformer encoder. Then we apply the memory augmenting method on its self-attention layer. We further propose an improved memory structure and make comparison experiments. Our contributions are as follows. First, we make further verification that self-attention layer is not necessary for low-level front layers and only a few self-attention layers added to high level can achieve competitive performance. Second, we apply augmented persistent memory to ASR model and we propose an improved variant of memory structure, which is more compact and competitive on recognition performance. All experiments are performed on a large-scale training data, ie., over 10,000hrs. 

\section{model architecture}
\label{sec:format}
A deep feed-forward sequential memory network \cite{dfsmn} provides a non-recurrent architecture to model long term dependency. Compared with self-attention network, FSMN's layer structure is much simpler and more focusing on a local range of neighbouring frames, while ignores the relative dependency in different positions of a sequence. A multi-head self-attention layer can model the relative dependency by gathering the information from the whole context in a sequence. In view of this, we explore the combination of FSMN layers and multi-head self-attention layers.  We conjecture that this proposed architecture can achieve a better trade-off between modeling efficiency and capturing the long term relative dependency.

\subsection{DFSMN}
DFSMN can conceptually be viewed as the standard feedforward fully connected neural networks augmented with some FIR-like filters.
The formulation of the FIR-like filters takes the following form:
\begin{equation}\label{equ:1}
\mathnormal{{h_{t}}}=\mathnormal{\sum_{i=0}^{N_{1}}a_{i} \cdot h_{t-i}} + \mathnormal{\sum_{j=1}^{N_{2}}b_{j} \cdot h_{t+j}}
 \end{equation}
where $N_{1}$ and $N_{2}$ denote the look-back and look-ahead order respectively. From equation 1, we can observe that the learnable FIR-like filters in DFSMNs can be used to encode long context information into fixed-size representation ($h_{t}$), which makes the DFSMN capture long-term dependency. However, the relative dependency in different positions is ignored in this architecture.

\subsection{SAN}
A self-attention network \cite{vaswani2017attention} has two sub-modules including a multi-head attention layer and a position-wise feedforward layer. In addition, dropout,
residual connection, and layer normalization are applied after both the self-attention and feed-forward layers. The computation process of self-attention is formulated as follows:
\begin{small}
\begin{equation}\label{equ:7}
\mathnormal{SelfAttn(Q,K,V)}=\mathnormal{softmax(\frac{QK}{\sqrt{d_{k}}})V}
 \end{equation}
 \end{small}
 \begin{small}
\begin{equation}\label{equ:7}
\mathnormal{MultiHead(Q,K,V)}=\mathnormal{Concat(head_{1},...,head_{h})}\mathnormal{W^{O}}
 \end{equation}
 \end{small}
  \begin{small}
 \begin{equation}\label{equ:7}
\mathnormal{head_{i}}=\mathnormal{SelfAttn(Q{W_{i}}^{Q},K{W_{i}}^{K},V{W_{i}}^{V})}
 \end{equation}
 \end{small}
where $W_{k}$, $W_{v}$ and $W_{q}$ are the key, value and query matrices of  size $d_{k}$ x $d$.  $d_{k} = {d}/{h}$ and h is the number of heads in the self-attention layer.  The multi-head attention performs by attending the information from different subspaces mapped by ${W_{i}}^{Q}$, ${W_{i}}^{K}$ and ${W_{i}}^{V}$. ${W^{O}}$ is the output weight matrix. It is clear that self-attention can explore the information from different representations at different positions. In other words, this architecture has the ability to model the relative dependency.

\subsection{DFSMN-SAN}
We propose the DFSMN-SAN model in which the multi-head self-attention layer (red block in Fig.1) is combined with DFSMN model.  
Similar to the combination of TDNN and SAN in \cite{povey2018time}, we argue that the combination of DFSMN and SAN can achieve a better trade-off between modeling efficiency and capturing the long-term relative dependency.  
Two types of the combination are empirically evaluated. The first is to simply stack all of the self-attention layers at the end of DFSMN model, and the second is to insert self-attention layer into DFSMN with an alternate style.
In our pilot experiments, we find that the latter consistently performs better. Therefore the combination with an alternate type is used in this paper, as shown in Fig. 1. After each 10 consecutive FSMN layers, a self-attention layer is inserted.

\begin{figure}[!tb]
\begin{minipage}[b]{1.0\linewidth}
  \centering
  \centerline{\includegraphics[width=8.2cm]{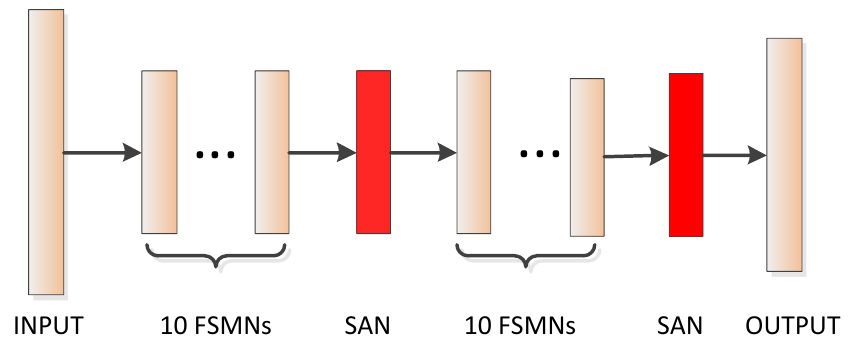}}
\end{minipage}

\caption{DFSMN-SAN model architecture}
\end{figure}

\section{AUGMENTING SELF-ATTENTION LAYERS WITH PERSISTENT MEMORY}
\label{sec:pagestyle}

In this section, we further propose to apply self-attention layer augmented with persistent memory into the DFSMN-SAN model. The motivation behind this is to investigate whether even more information beyond the whole utterance level can be exploited and beneficial.
These memory vectors are random initialized from the beginning of training and updated with whole training corpus. We believe that these memory vectors can learn and store some global information useful for the ASR task. Here we propose a new memory structure different with \cite{MEM}. The two different types of memory structures are described as follows.

\subsection{Key-Value Memory structure}

\label{sec:typestyle}
\begin{figure*}[!tb]
\begin{minipage}[b]{1.0\linewidth}
  \centering
  \centerline{\includegraphics[width=12.5cm]{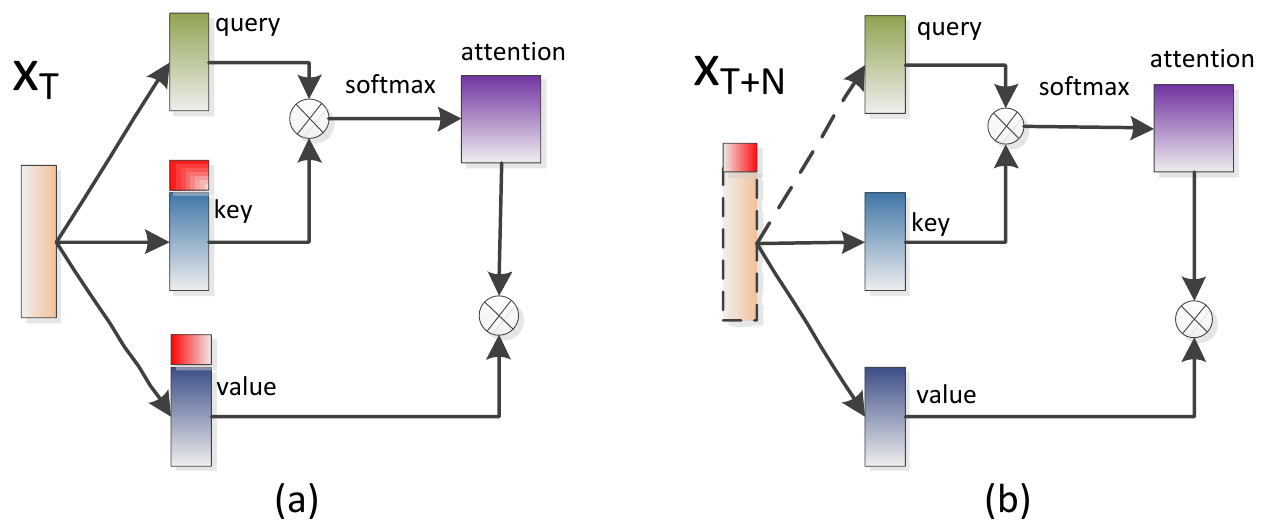}}
\end{minipage}
\vspace{-15pt}
\caption{In Fig.2(a), the persistent memory vectors are concatenated to key-value vectors. In Fig.2(b), the persistent memory vectors are directly concatenated to input vectors. $X_{T+N}$ denotes the input vectors $X_{T}$ augmented with persistent vectors. The key-value vectors are generated from $X_{T+N}$ and query vectors are generated from $X_{T}$ (dotted line in Fig.2(b)).  We represent the memory vectors with a red block for both models. Here, we represent both models in the case of a single head. In our experiments, both models have multiple heads.}
\vspace{-15pt}
\end{figure*}

Fig.2(a) shows the self-attention layer augmented with persistent memory vectors proposed by \cite{MEM}. More precisely, these persistent memory vectors are a set of N pairs key-value vectors, which are stacked in two $d_{k}$ x $N$-dimensional matrices $M_{k}$ and $M_{v}$.  

These persistent memory vectors are simply concatenated to the pool of key-value vectors:
\begin{equation}\label{equ:7}
\mathnormal{K_{m}}=\mathnormal{Concat([W_{k}x_{1},...,W_{k}x_{T}],M_{k})}
 \end{equation}
 \begin{equation}\label{equ:8}
 \mathnormal{V_{m}}=\mathnormal{Concat([W_{v}x_{1},...,W_{v}x_{T}],M_{v})}
 \end{equation}
\begin{equation}\label{equ:9}
 \mathnormal{SelfAttn(Q,K_{m},V_{m})}=\mathnormal{softmax(\frac{QK_{m}}{\sqrt{d_{k}}})V_{m}}
 \end{equation}
where the position encoding corresponding to a persistent vector is equal to zero. $\mathnormal{X_{T}=[x_{1},...,x_{T}]}$. $K_{m}$ denotes the concatenation of the key vectors with the corresponding N persistent vectors.  
Similar to general multi-head self-attention layers, the persistent memory vectors are split into multiple heads and not shared between heads. 

\subsection{Input-Embedding Memory structure}
In this paper, we propose a new type of memory structure which is shown in Fig.2(b). Different from key-value memory structure, these persistent memory vectors are directly concatenated to the input vectors:
\begin{small}
\begin{equation}\label{equ:10}
\mathnormal{K_{m}}=\mathnormal{Concat([W_{k}x_{1},...,W_{k}x_{T}],[W_{k}M_{1},...,W_{k}M_{N}])}
 \end{equation}
 \end{small}
 \begin{small}
 \begin{equation}\label{equ:11}
 \mathnormal{V_{m}}=\mathnormal{Concat([W_{v}x_{1},...,W_{v}x_{T}],[W_{v}M_{1},...,W_{v}M_{N}])}
 \end{equation}
  \end{small}
where $[M_{1},...,M_{N}]$ denote the persistent memory vectors. Obviously, $K_{m}$ and $V_{m}$ share the same persistent memory vectors. That is to say, we have fewer parameters compared with key-value memory structure.

\section{EXPERIMENTAL SETUP}
\label{sec:format}
\vspace{-5pt}
\subsection{Training setup}
\vspace{-5pt}
 The feature vectors used in all the experiments are 40-dimensional log-Mel filterbank energy features appended with the first-order and the second-order derivatives. Log-mel filterbank energy features are computed with a 25ms window and shifted every 10ms. We stack 8 consecutive frames and subsample the input frames with 3. A global mean and variance normalization is applied for each frame. All the experiments are based on the CTC learning framework. We use the CI-syllable-based acoustic modeling method \cite{syllable} for CTC learning. The target labels of CTC learning are defined to include 1394 Mandarin syllables, 39 English phones, and a blank. Character error rate results are measured on the test sets. We use a pruned, first pass, 5-gram language model. All the systems use a vocabulary that consists of millions of words. Decoding is performed with a beam search algorithm by using the weighted finite-state transducers (WFSTs).

\vspace{-12pt}
\subsection{Datasets}
\vspace{-5pt}
Our training corpus is mixed data sets collected from several different application domains, all in Mandarin.   In order to improve system robustness,  a set of simulated room impulse responses (RIRs) are created with different rectangular room sizes, speaker positions, and microphone positions, as proposed in \cite{far2}. Together they contain a total of 10k hours speech. 

To evaluate the performance of our proposed method, we report performance on 3 types of test sets which consist of hand-transcribed anonymized utterances extracted from reading speech (1001 utterances), conversation speech (1665 utterances) and spontaneous speech (2952 utterances).  We refer them as Read, Chat, and Spon respectively. In addition, to provide a  public benchmark, we also use AISHELL-2 development set (2500 utterances) recorded by high fidelity microphone as the test set.


\begin{table}
\caption{\textit{Results of the different model architectures. } }
\vspace{1pt}
\label{tab:1}
\begin{center}
\begin{tabular}{|c|c|c|c|c|c|}
\hline
\multirow{2}*{\small{Model}} & \multirow{2}*{\small{Size}} & \multicolumn{4}{c|}{\small{Test set}} \\
\cline{3-6}&  & \small{Read} & \small{Chat} & \small{Spon} &  \small{AISHELL}  \\
\hline
\small{DFSMN}    & 131M     & 3.19 & 31.59 & 32.82  & 5.78 \\

\small{SAN}      & 141M    & 2.66  & 30.29 & 30.40 & 5.24 \\

\small{DFSMN-SAN}  & 143M  & \bf{2.09}  & \bf{28.56} & \bf{28.70} & \bf{4.95}\\

 \hline
\end{tabular}
\end{center}
\end{table}
\subsection{Acoustic Model}
For the first experiment, we present our work with DFSMN, self-attention and DFSMN-SAN model. The DFSMN system uses 30 DFSMN components of 1024 hidden units, each with a projection layer of 512 units. 
The self-attention model contains 10 multi-head self-attention sublayers with a comparable size with 30 DFSMN model. We set the model dimension $d = 512$ and the number of heads $h = 8$. The DFSMN-SAN model consists of 30 DFSMN components and 3 multi-head self-attention sublayers. 

For the second experiment, we improve the DFSMN-SAN by augmenting the self-attention sublayer with persistent memory. We set the number of heads to 8 for key-value memory vectors. The position embedding is shared across all the heads.

For stable CTC learning, we clip gradients to [-1.0, 1.0].
We use the Kaldi \cite{kaldi} toolkit to train models and all models are trained in a distributed manner using BMUF \cite{BMUF1} optimization with 8 Tesla P40 GPUs.

\section{Experimental Results}
\subsection{Model Architecture Experiments}
In this section, we compare different variations of our acoustic models. Table 1 shows the performance comparison of 3 different acoustic models. Line 2 presents the results of the baseline system (30 layers FSMN).  Line 3 presents the results of self-attention model of 10 layers. The last line presents the results of DFSMN-SAN model. The results show that self-attention model performs better than DFSMN model with comparable model size. As expected, the DFSMN-SAN model which combines the dfsmn layers with self-attention layers performs best compared with other models. This indicates that the relative dependency learned by self attention layers can improve the system performance while self-attention mechanism is not necessary for low-level front layers, and only adding a few self-attention layers to high level can achieve competitive performance. Notably, the DFSMN-SAN model achieves up to 34.4\% relative CER improvement over the baseline model on the Read test set.
\label{sec:print}

\subsection{Augmenting Memory Experiments}

Table 2 shows the experimental results of models augmented with key-value memory structure. The different number of persistent memory vectors $N$ was examined and the performance seems not to vary too much as $N$ set to be 64, 128 and 256. The key-value memory structure achieves 9.5\% relative CER improvement over the baseline model on the AISHELL test set.

Table 3 shows the models augmented with input-embedding memory structure also significantly outperform the baseline system. 
Both two types of augmented memory introduce very modest increase in model size. The proposed input-embedding memory structure has achieved competitive performance to the key-value memory structure with fewer parameters.
\vspace{-10pt}

\begin{table}
\caption{\textit{Results of the different number of persistent vectors (N) for key-value memory structure. } }
\vspace{-2pt}
\label{tab:1}

\begin{center}
\begin{tabular}{|c|c|c|c|c|c|}
\hline
\multirow{2}*{\small{Model}} & \multirow{2}*{\small{Size}} & \multicolumn{4}{c|}{\small{Test set}} \\
\cline{3-6}&  & \small{Read}  & \small{Chat} & \small{Spon} &  \small{AISHELL} \\
\hline
\small{Baseline}      &143M    & 2.09 & 28.56 & 28.70  & 4.95  \\

\small{N=64}          &144M & \bf{1.93}  & 27.16 & \bf{25.41} & 4.48 \\

\small{N=128}     &145M& 1.95  & 27.23 & 25.45 & \bf{4.47}\\
\small{N=256}     &146M& 1.96 & \bf{27.08} & 25.58  & 4.52\\

 \hline
\end{tabular}
\end{center}
\vspace{-20pt}
\end{table}

\begin{table}
\caption{\textit{Results of the different number of persistent vectors (N) for input-embedding memory structure.} }
\vspace{-2pt}
\label{tab:1}

\begin{center}
\begin{tabular}{|c|c|c|c|c|c|}
\hline
\multirow{2}*{\small{Model}} & \multirow{2}*{\small{Size}} & \multicolumn{4}{c|}{\small{Test set}} \\
\cline{3-6}&  & \small{Read} & \small{Chat} & \small{Spon} &  \small{AISHELL}   \\
\hline
\small{Baseline}        &143M  & 2.09  & 28.56 & 28.70 & 4.95  \\

\small{N=64}           &143M& \bf{1.95}  & 27.35 & \bf{25.72} & \bf{4.41} \\

\small{N=128}     &144M& 1.99  & 27.30 & 25.86 & 4.43 \\
\small{N=256}     &145M& 2.02 & \bf{26.89} & 26.13 & 4.50 \\

 \hline
\end{tabular}
\end{center}
\vspace{-20pt}
\end{table}

\section{CONCLUSION}
\label{sec:page}
\vspace{-8pt}
In this work, we first explore a variant model architecture that combines DFSMN layer with  self-attention layer. 
By adding  multi-head self-attention layer into DFSMN with an alternate type, consistent improvement can be obtained both over the original DFSMN model and the SAN model with pure self-attention layers. 
More importantly, we apply self-attention layer with persistent memory vectors into DFSMN-SAN model. Two types of augmented memory methods are evaluated and both provide further improvements in CER.  
Notably, we find that our proposed input-embedding memory structure can achieve comparable performance with the key-value memory structure with fewer parameters. 
To make experimental results more convincing, all experiments in this paper are performed on large data sets and evaluated four different individual test sets. 
Future work includes making visualization analysis on the memory vectors to get better understanding of what they have learned.



\vfill\pagebreak

\bibliographystyle{IEEEbib}
\bibliography{strings,refs}

\end{document}